# A new approach to instanton calculations in the O(3) nonlinear $\sigma$-model

Jeroen Snippe[*] and Pierre van Baal

Instituut-Lorentz for Theoretical Physics, University of Leiden,
PO Box 9506, NL-2300 RA Leiden, The Netherlands

We construct all instantons in the O(3) $\sigma$-model on a cylindrical space-time, known not to exist on a finite time interval. The scale parameter, $\rho$, is related to the boundary condition in time. This may cure the $\rho \to 0$ divergent instanton gas, through a proper inclusion of in and out states in the path integral.

## 1. Introduction

Traditionally, semiclassical calculations focus on instanton contributions to

$$Z(T) = \int \mathcal{D}\phi \exp\{-S[\phi]\} = \text{Tr} \exp\{-\mathcal{H}T\} \ , \quad (1)$$

Tr being the trace over all quantum states and $T$ being the time interval. The physically interesting objects, however, are long distance – low energy quantities. For these one only needs to know $\langle n| \exp\{-\mathcal{H}T\}|n\rangle$, where $|n\rangle$ is a low energetic state.

If $|n\rangle$ corresponds to the ground state, formally $\langle n| \exp\{-\mathcal{H}T\}|n\rangle$ may be derived from $Z(T)$ by studying the limit $T \to \infty$. Technically this is difficult and involves as an intermediate step the dilute gas approximation [1]. What is more, analytical calculations have often been done on spherical space-times, both implicitly [2] and explicitly ([3] and references therein). In such cases a Hamiltonian interpretation is only possible for the radius $R \to \infty$. However, in asymptotically free models like 4-D SU($N$) and 2-D O(3), taking $R \to \infty$ makes the semiclassical approximation unreliable.

Therefore we study the O(3) $\sigma$-model on the cylinder, $S^1 \times \mathbb{R}$. Not only does this space-time admit a Hamiltonian formalism, but also the coupling constant can be tuned arbitrarily through the length of the spatial $S^1$. Furthermore it allows for a lattice formulation, important for comparison to simulations. Finally, as will be shown below, it admits instantons. This is unlike $S^1 \times [-\frac{1}{2}T, \frac{1}{2}T]$, which only admits instantons [4] with topological charge two or more. One can easily argue that (for large $T$), nevertheless, approximate solutions of topological charge one are the most important non-trivial configurations. But as they are not exact solutions, it is not quite clear how these can be incorporated in conventional semiclassical calculations.

In section 2 we will briefly sketch the construction and physical interpretation of instantons on $S^1 \times \mathbb{R}$ found by one of us. For details see ref. [5]. This reference also establishes the exact relation between instantons, sphalerons and the vacuum valley. This would be important for an alternative non-perturbative approach like in [6]. In section 3 we will comment on aspects of the semiclassical calculation, which is in progress.

## 2. Instantons

The action of the O(3) $\sigma$-model is well-known and reads

$$S = \frac{1}{2f} \int_{T^1 \times \mathbb{R}} d^2x \ |\partial_\mu \mathbf{n}(x)|^2, \quad \mathbf{n}(x) \in S^2. \quad (2)$$

In terms of complex coordinates $z = (x_1 + ix_2)$, $\bar{z} = (x_1 - ix_2)$, the (charge one) instanton solutions on $\mathbb{R}^2$ are rational functions [7]

$$u_{\mathbb{R}^2} = -\frac{c+dz}{a+bz} \quad (a,b,c,d \in \mathbb{C}, ad-bc=1), \quad (3)$$

with $u$ the complex stereographic parametrization of $S^2$ [7,4]. Since the action, eq. (2), is conformally invariant, eq. (3) leads to the following in-

---
[*]Based on the talk presented by the first author at Lat'94 (Bielefeld, Sept. 1994)

stanton solutions on $S^1 \times \mathbb{R}$:

$$u = -\frac{c + de^z}{a + be^z}. \qquad (4)$$

where $z$ is now expressed in units of $L$, the spatial volume. It is not difficult to prove [5] that this is again the most general charge one instanton. The moduli space $\{a, b, c, d \in \mathbb{C}, ad - bc = 1\}$ has therefore six real dimensions.

In order to obtain a more physical parametrization note that, with $T \to \infty$, an instanton always has to start and end in the vacuum valley, to keep its action finite. In our case this is the space of constant fields $S^1 \to S^2$, i.e. $S^2$ itself. Indeed, eq. (4) gives

$$u_\pm \equiv \lim_{x_1 \to \pm\infty} u(x_1, x_2) = \begin{cases} -d/b \\ -c/a. \end{cases} \qquad (5)$$

The remaining two parameters correspond to the instanton position.

We can also use a parametrization given by the orbit under the symmetry group of the model. This parametrization consists of three SO(3) parameters and two translational (space and time) parameters. The remaining sixth parameter can be easily interpreted as a scale parameter [5]. The exact correspondence to $u_\pm$ is

$$\rho = d(u_+, u_-), \qquad (6)$$

where $d$ stands for the geodesic distance on $S^2$. Maximal $\rho$ corresponds to maximal (antipodal) distance. The value $\rho = 0$ is reached in the limit $u_+ = u_-$, which is not a regular point of the moduli space since it violates the condition $ad - bc = 1$ (it would formally make $u$ in eq. (4) constant, and of topological charge zero).

## 3. Semiclassical aspects and outlook

The hope is that due to the new interpretation of the scale parameter in terms of degenerate classical vacua the semiclassical $\rho \to 0$ behavior can be calculated in a more controlled fashion. By working in a finite spatial volume, one avoids possible interference with infrared divergences. The cylindrical geometry admits a Hamiltonian framework. Both of these ingredients we consider important. It should be noted that recent numerical simulations [8] show a rapid increase of the instanton density for decreasing $\rho$. In particular simulations with the perfect action [9] for the O(3) model, where lattice artefacts are under control, give a topological susceptibility that fails to scale up to correlation lengths of 60 lattice spacings at fixed physical volume [10]. As was noted in ref. [10], it seems unavoidable that this is caused by an increasing contribution of small instantons for decreasing lattice spacing, implying a divergent topological susceptibility in the continuum limit. Consequently the second order derivative of the free energy with respect to $\theta$ at $\theta = 0$ diverges. We hope that our analysis will be powerful enough to address this issue in an analytic setting.

It is convenient, for practical purposes only, to use the equivalence of the O(3) model to the $CP^1$ model [11]. The U(1) local gauge symmetry of the latter allows a simpler description of instanton contributions in terms of tunneling between different vacua (or vacuum valleys), instead of separating these contributions in the path integral through their winding number. On the physical configuration space (i.e. after dividing out the local gauge symmetry in the $CP^1$ model), the two pictures are equivalent.

We propose to use the Green's function techniques that were developed to split the path integral into contributions coming from classically allowed and forbidden regions, each amenable to its own approximate method of calculation. This method is known as the Path Decomposition Expansion (PDX), originally developed for applications in condensed matter physics [12], but successfully used in the semiclassical tunneling calculations [13] for gauge theories on $T^3 \times \mathbb{R}$. Also there the vacuum valley was non-trivial, except that no scale parameter was involved. The relevant features of the PDX were reviewed in ref. [14].

In the PDX one splits configuration space into regions around the classical vacua and their complement which give rise to a restricted Green's function $G^{(i)}$ that vanishes on the boundary $\Sigma_i$ of these regions. In the present case the vacua form valleys which have to be fully contained inside $\Sigma_i$. Up to this surface the Green's function is described well enough by perturbation theory, its relevant contribution being determined by the perturbative wave functional and energy of the state for which one wishes to calculate the shift in energy due to the tunneling processes. The position of the surfaces $\Sigma_i$ can best be specified by requiring the classical turning points, determined from the perturbative energy, to lie well within $\Sigma_i$. The perturbative wave functional is in lowest order simply given by harmonic functions for the non-zero momentum modes and constant functions for the zero momentum modes. In particular the non-Gaussian behavior in the latter needs to be treated with care.

The tunneling contributions mentioned in the previous paragraph are captured by a transition Green's function $G^{(tr)}$ which also has vanishing boundary conditions on each of the surfaces $\Sigma_i$. In the path integral representation of this Green's function only those paths contribute that never enter the perturbative regions bounded by $\Sigma_i$. These paths are entirely within the classically forbidden region and can be computed using semiclassical techniques.

It can be rigorously shown that the shift in energy due to the tunneling is determined by the following matrix element [14]:

$$<\Psi^{(i)}|[\Sigma_i]G^{(tr)}[\Sigma_j]|\Psi^{(j)}>, \quad (i \neq j). \tag{7}$$

In this expression $\Psi^{(i)}$ is the perturbative wave functional on the surface $\Sigma_i$, while the operator $[\Sigma_i]$ involves a two-side derivative with respect to the normal of, and an integration over, this surface. Work is in progress to compute to sufficient accuracy the Green's function $G^{(tr)}$, incorporating correctly the ultraviolet divergencies due to the Gaussian fluctuations, and finally perform the surface integrations. In this, only a few modes have to be treated in a non-Gaussian approximation, such that it is expected the calculations can be performed without too much difficulty.

Crucial is that in the semiclassical part of these calculations, described by the Green's function $G^{(tr)}$, we never connect identical points in the vacuum valley. This is because $G^{(tr)}$ involves determining the classical solutions for paths that start and end on different $\Sigma_i$. The end points therefore never get close to the vacuum valleys. It remains to be investigated that this indeed regulates the $\rho \to 0$ problem, and that the problem does not reappear elsewhere [15]. In any case the PDX method does not rely on a dilute gas picture and takes initial and final states properly into account. In this way we hope to be able to address the issue of defining the $\theta$ dependence of the groundstate energy, and to arrive at a sufficiently clear physical picture of the behavior of the model in this respect.

One of us (PvB) is grateful to Rudolf Burkhalter, Peter Hasenfratz, Martin Lüscher, Ferenc Niedermayer and Peter Weisz for discussions.